\documentclass[12pt,preprint]{aastex}
\newcommand{\beq}{\begin{equation}}
\newcommand{\beqa}{\begin{eqnarray}}
\newcommand{\eeq}{\end{equation}}
\newcommand{\eeqa}{\end{eqnarray}}

\newcommand{\siml}{\lesssim}
\shorttitle{Ultra Fast Self-Compton Cooling}
\shortauthors{Ioka}
\begin{document}
\title{
Ultra Fast Self-Compton Cooling
}
\author{
Kunihito Ioka
}
\affil{Department of Earth and Space Science,
Osaka University, Toyonaka 560-0043, Japan}
\email{ioka@vega.ess.sci.osaka-u.ac.jp}

\begin{abstract}
We investigate the synchrotron self-Compton process 
in a planar shell taking the shock structure into account.
We find that the energy density of the seed photons could 
deviate from the one-zone
estimate by order of unity depending on the shock velocity
and the electron cooling time.
We also find that as the electron cooling becomes faster,
the seed photons are increased more,
so that the inverse Compton cooling becomes more efficient.
This ``ultra'' fast cooling may work in such as gamma-ray bursts, blazars and
microquasars.
\end{abstract}

\keywords{galaxies: jets --- gamma rays: bursts --- gamma rays: theory
--- radiation mechanisms: non-thermal --- shock waves}

\section{INTRODUCTION}\label{sec:intro}
Relativistic shocks often arise in astrophysics
when a faster flow hits upon a slower one
such as in gamma-ray bursts (GRBs) (e.g., Piran 1999),
blazars (e.g., Inoue \& Takahara 1996; Kino, Takahara \& Kusunose 2002;
Rees 1978), 
microquasars (e.g., Mirabel \& Rodr\'{\i}guez 1999;
Levinson \& Waxman 2001; Kaiser, Sunyaev \& Spruit 2000) and so on.
In the relativistic shocks, the kinetic energy of the flow
turns into the internal one,
and some fraction of the internal energy is distributed to
electrons and magnetic fields.
The electrons are accelerated in the shock front
while the magnetic fields are amplified by the shock.
Under these conditions, the accelerated electrons radiate
nonthermal emission, such as synchrotron and inverse Compton (IC) emission.
In this paper we will consider the synchrotron
self-Compton (SSC) process, in which the seed photons for the IC
emission are the synchrotron photons.

The intensity of the IC emission is proportional to the 
energy density of the seed photons (Rybicki \& Lightman 1979).
So far the energy density of the seed photons has been estimated
from the one-zone argument
(e.g., Inoue \& Takahara 1996; Sari, Narayan \& Piran 1996; 
Sari \& Esin 2001).
In the one-zone model we assume that the seed photons are distributed
uniformly over space.
However the one-zone approximation is too crude at times.
In fact the shock structure,
such as the cooling layer (e.g., Granot, Piran \& Sari 2000)
or the magnetized layer (e.g., Rossi \& Rees 2002),
could bring new features.
Also for the seed photons, the shock structure may be important 
since the uniformity is violated near the shock front.
At least we should evaluate the validity of the one-zone approximation
quantitatively taking the shock structure into account.

In this paper, we will investigate the effects of the shock structure
on the energy density of the seed photons in the SSC process.
Especially we will concentrate on the geometrical effects.
We will find that for some parameters
the energy density of the seed photons could deviate from
the one-zone estimate by order of unity.

\section{PLANAR MODEL}
Let us consider the following simple model.
Our model is not so sophisticated as the realistic ones.
However simplicity should be helpful in elucidating the main features.

\begin{figure}
\plotone{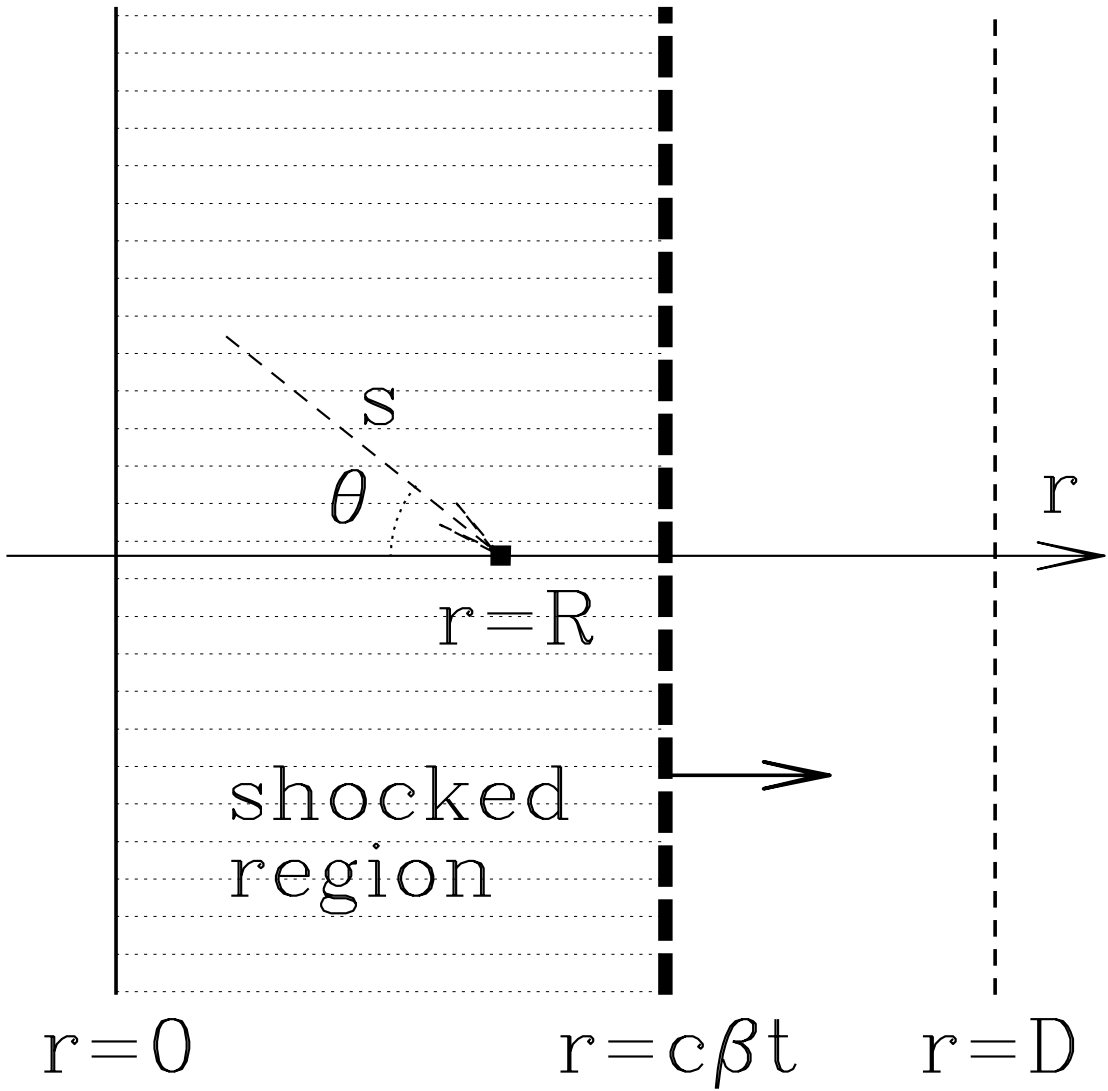}
\caption{Our simple model is shown.
We consider an optically thin uniform shell with a thickness $D$
and a shock propagating with a velocity $c \beta=c (1-\gamma^{-2})^{1/2}$
measured in the shocked fluid frame.
Electrons are accelerated just behind the shock,
and then they cool radiating synchrotron and Compton emission.
The motion of the electrons is neglected in the shocked fluid frame.
}\label{fig:model}
\end{figure}

We consider an optically thin uniform shell with a thickness $D$
and a shock propagating with a velocity $c \beta=c (1-\gamma^{-2})^{1/2}$
measured in the shocked fluid frame (see Fig.~\ref{fig:model}).
Electrons are accelerated just behind the shock
and the motion of the electrons is negligible in the shocked fluid frame,
if the electron acceleration time
is shorter than other timescales.
To extract the geometrical effects,
we assume that all particles and magnetic fields are isotropic
in the shocked fluid frame.
We consider only single scattering 
assuming that the higher order IC is suppressed
by the Klein-Nishina effect.

Note that if the shock is strong and the shocked fluid is extremely hot,
the shock velocity measured in the shocked fluid frame is given by 
$\beta=[(\gamma_{r}-1)/(\gamma_{r}+1)]^{1/2}/3$
(Blandford \& Mckee 1976),
where $\gamma_{r}\sim (\gamma_{f}/\gamma_{s}+\gamma_{s}/\gamma_{f})/2$ is
the relative Lorentz factor 
between the faster fluid with the Lorentz factor $\gamma_{f}$ and 
the slower fluid with $\gamma_{s}$.
When $\gamma_{f}/\gamma_{s}\sim 2$ and $\gamma_{f}/\gamma_{s}\gg 1$, 
$\beta\sim 0.1$ and $\beta=1/3$, respectively.
Therefore we will consider the cases $\beta=0.1$ and $\beta=1/3$.

The energy density of the seed photons at a point $r=R$ is given by
the integration of all synchrotron radiation from the accelerated electrons as
\beqa
U_{\gamma}\left(t,r=R\right)=\frac{2 \pi}{c}
\int d\mu ds \frac{1}{4\pi} P\left(t-\frac{s}{c},r\right),
\label{eq:ugsum}
\eeqa
where $P(t,r) \ {\rm erg}\ {\rm s}^{-1}{\rm cm}^{-3}$
is the synchrotron power at a time $t$ and a position $r$,
$\mu=\cos \theta$ and $s>0$ (see Fig.~\ref{fig:model}).
Note that the retarded time $t-s/c$ is used in equation (\ref{eq:ugsum}).
For simplicity, we consider the monoenergetic injection of the electrons.
Since we concentrate on the total energy density of the seed photons,
we can regard these monoenergetic electrons as the
electrons that contribute most of the radiation energy 
even for the power-law injection case.
To extract the essence, we adopt
the following simple form as the synchrotron power,
\beqa
P(t,r)=A\left[H\left(t-\frac{r}{c\beta}\right)
-H\left(t-\frac{r}{c\beta}-\Delta t_{cool}\right)\right],
\label{eq:psyn}
\eeqa
which means that electrons radiate 
with a constant power
$A \ {\rm erg}\ {\rm s}^{-1} \ {\rm cm}^{-3}$
for a duration $\Delta t_{cool}\ {\rm s}$ since the shock has passed.
This is sufficient for the following argument,
since the synchrotron power is proportional to the square of the electron
Lorentz factor (Rybicki \& Lightman 1979) and hence
the cooled electrons have a small contribution to 
the total energy density of the seed photons.

Equation (\ref{eq:psyn}) makes it possible to
integrate equation (\ref{eq:ugsum}) analytically.
As shown in \S~\ref{sec:app}, during the electron emission,
\beqa
0<\tilde t \equiv t-\frac{R}{c\beta}<\Delta t_{cool},
\label{eq:temit}
\eeqa
the energy density in equation (\ref{eq:ugsum}) is given by
\beqa
U_{\gamma}(t,R)
=\frac{A}{2}\Biggl[
\beta (\Delta t_{cool}-\tilde t)
\ln \frac{(1-\beta)(t-\Delta t_{cool})}
{\beta (\Delta t_{cool}-\tilde t)}
+\frac{R}{c}\ln \frac{t}{t-\Delta t_{cool}}
+\beta \tilde t \ln \frac{(1+\beta)t}{\beta \tilde t}
\Biggr],
\label{eq:ugint1}
\eeqa
if
\beqa
\tilde t>\Delta t_{cool}-\frac{(1-\beta)R}{c\beta},
\label{eq:t3}
\eeqa
and 
\beqa
\tilde t < \frac{(1+\beta)(D-R)}{c\beta},
\label{eq:td}
\eeqa
are satisfied.
If equations (\ref{eq:temit}) and (\ref{eq:td}) 
are satisfied but equation (\ref{eq:t3}) is not,
we have
\beqa
U_{\gamma}(t,R)
=\frac{A}{2}\Biggl[
\frac{R}{c}\ln \frac{ct}{R}
+\beta \tilde t \ln \frac{(1+\beta)t}{\beta \tilde t}
\Biggr].
\label{eq:ugint2}
\eeqa

\section{ENERGY DENSITY OF SEED PHOTONS}
Let us consider the energy density of the seed photons
when the shock is in the region $0<r<D$.
The shock crossing time is given by $\Delta t_{dyn}=D/c\beta$.
We call the case $f\equiv \Delta t_{dyn}/\Delta t_{cool}> 1$ 
fast cooling, and $f<1$ slow cooing.

First, let us consider the one-zone model.
In fast cooing, electrons emit almost all energy before
the shock crosses the shell, so that
the energy density of the seed photons is given by
\beqa
U_{\gamma}^{1zone}=\int_{0}^{\Delta t_{cool}}
d\tilde t P(t,r)=A \Delta t_{cool},
\label{eq:1zone}
\eeqa
with equation (\ref{eq:psyn}).
In slow cooing, electrons do not emit
all the energy within $\Delta t_{dyn}$.
The electrons radiate with a power $A$
for a duration $\Delta t_{dyn}$,
so that 
\beqa
U_{\gamma}^{1zone}=A \Delta t_{dyn}=\frac{A D}{c\beta}.
\eeqa

In the planar model, the energy density of the seed photons 
is given by equations (\ref{eq:ugint1}) and (\ref{eq:ugint2}).
To compare the planar model with the one-zone model,
we take the time and position average of the energy density,
$\langle U_{\gamma}(t,r) \rangle
\equiv \int_{0}^{D} (dr/D) \int_{0}^{\Delta t} 
(d\tilde t/\Delta t) U_{\gamma}(t,r)$
where $\Delta t \equiv \min[\Delta t_{cool},(D-r)/c\beta]$.
In the fast cooling limit $f\gg 1$, 
we have
\beqa
\langle U_{\gamma}(t,r) \rangle
=\frac{\beta A \Delta t_{cool}}{2} \left[
\frac{1}{2}+\ln \frac{f}{\beta \gamma}\right].
\label{eq:ugfast}
\eeqa
Note that the logarithmic term originates 
from the integral of $\sim 1/\mu$, i.e., the geometrical effect
(see \S~\ref{sec:app}).
In the slow cooling limit $f \ll 1$,
we have
\beqa
\langle U_{\gamma}(t,r) \rangle=
\frac{A D}{24 c}\left[\pi^2-6+3 \ln \frac{1+\beta}{\beta^3}\right].
\label{eq:ugslow}
\eeqa

In Fig.~\ref{fig:ave} we show the ratio of the energy density
of the seed photons to the one-zone estimate 
$\langle U_{\gamma}(t,r) \rangle /U_{\gamma}^{1zone}$
for $\beta=1/3$ and $\beta=0.1$.
The analytical approximations in equations 
(\ref{eq:ugfast}) and (\ref{eq:ugslow}) are also shown by dashed lines.
We see that the analytical approximations are quite good.

From Fig.~\ref{fig:ave}, we can find the following features.
First, for some parameters, e.g., $\beta=0.1$ and $f\siml 1$,
the energy density of the seed photons deviates from the one-zone estimate
by order of unity.
Second, the dependence of the energy density on the shock velocity $c \beta$
differs between in the planar model and the one-zone model.
Finally, in fast cooling the energy density of the seed photons
increases as the ratio $f$ increases,
because of the logarithmic term in equation (\ref{eq:ugfast}).
This is an interesting new feature due to the geometrical effect.
Increasing the seed photons enhances the IC emission
and hence the IC cooling.
Therefore this mechanism may be termed ``ultra'' fast cooling.
The difference between $f=1$ and $f\sim 10^{6}$ is about order of unity
when $A \Delta t_{cool}={\rm const}$.
Surprisingly the energy density of the seed photons
can be larger than that of the electrons which supply the seed photons.

\begin{figure}
\plotone{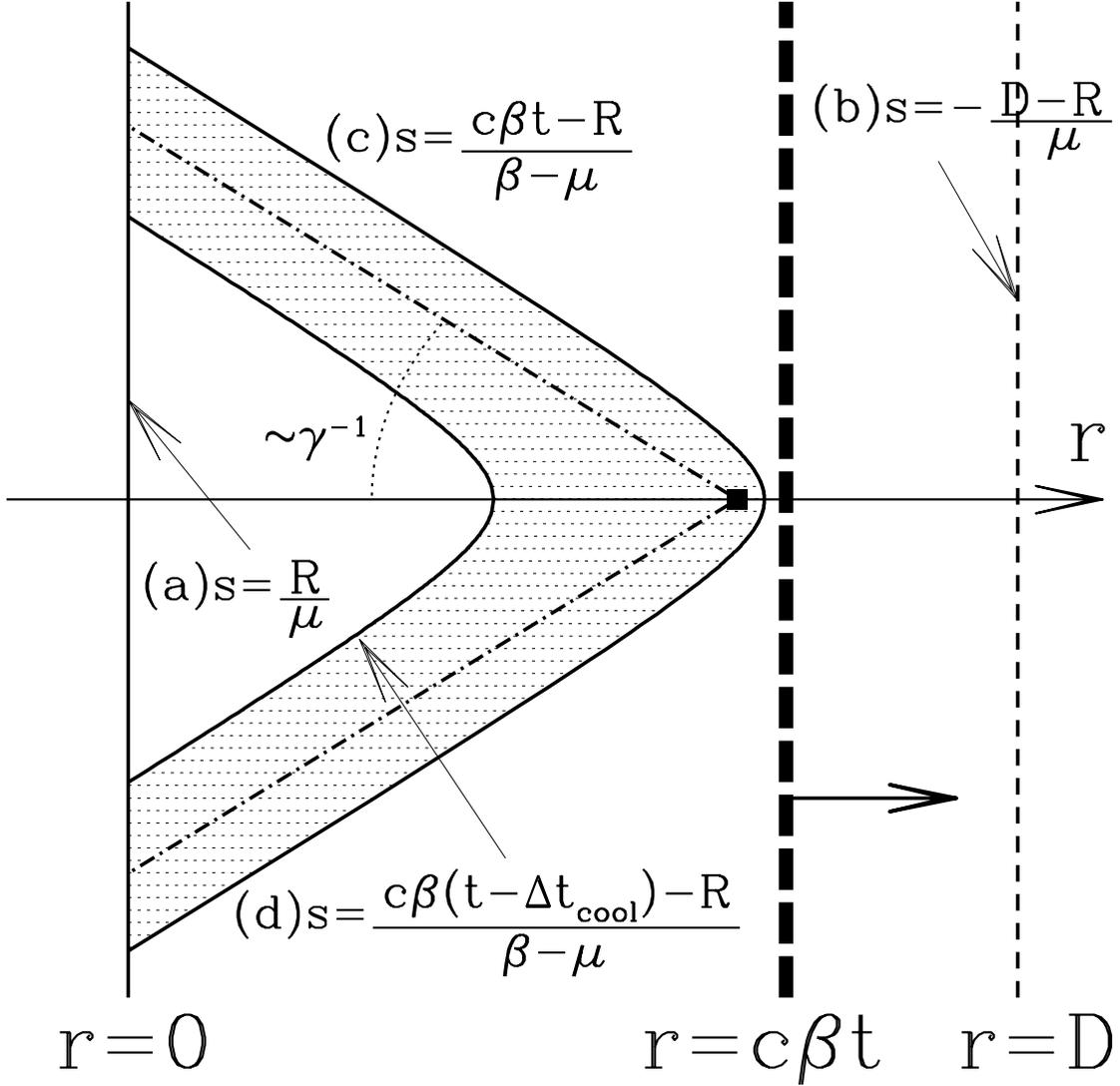}
\caption{The region from which photons come 
to the square point is shaded (see also Fig.~\ref{fig:model}),
where we assume that equations (\ref{eq:temit}), (\ref{eq:t3}) 
and (\ref{eq:td}) are satisfied.
This causal region is bounded by four constraints
as in Fig.~\ref{fig:region}, (a) $s=R/\mu$, (b) $s=-(D-R)/\mu$,
(c) $s=(c\beta t-R)/(\beta-\mu)$,
and (d) $s=[c\beta (t-\Delta t_{cool})-R]/(\beta-\mu)$.
The boundaries (c) and (d)
are the hyperboloid of two sheets.
The boundary (b) is not relevant to this figure.
Approximately, photons come from the direction 
$\tan \theta = \beta^{-1}\gamma^{-1}$,
which is shown by the dot-dashed lines.
}\label{fig:photon}
\end{figure}

To make the physical situation clear,
the region from which photons come is shade in Fig.~\ref{fig:photon}.
Here we assume that equations (\ref{eq:temit}), (\ref{eq:t3}) and (\ref{eq:td})
are satisfied.
This causal region is bounded by four constraints
as in Fig.~\ref{fig:region} (see \S~\ref{sec:app}), 
(a) $s=R/\mu$, (b) $s=-(D-R)/\mu$,
(c) $s=(c\beta t-R)/(\beta-\mu)$
and (d) $s=[c\beta (t-\Delta t_{cool})-R]/(\beta-\mu)$.
The boundary (c) represents the hyperboloid of two sheets,
\beqa
\frac{x^2+y^2}{\gamma^2 c^2 \beta^2 \tilde t^2}
-\frac{(r-R-\gamma^2 c \beta \tilde t)^2}
{\gamma^4 c^2 \beta^4 \tilde t^2}=-1,
\label{eq:hyper}
\eeqa
where $x^2+y^2=s^2(1-\mu^2)$.
The boundary (d) is also described by equation (\ref{eq:hyper})
with $\tilde t$ replaced by $\tilde t-\Delta t_{cool}$.
From equation (\ref{eq:hyper}) we can see that
the photons approximately come from the direction 
$\sim \tan \theta =\beta^{-1}\gamma^{-1}$.
This is physically reasonable
since the cooling layer for fast cooling is very close to the shock front and
the shock front has the same velocity 
towards the $r$-axis as the photons traveling in the direction of
$\tan \theta = \beta^{-1} \gamma^{-1}$.
In a sense, photons are accumulated around the shock front
as the shock sweeps.
The approximation of the planar shell
is valid when the size of the shell is larger than
$\sim \beta^{-1} \gamma^{-1} D$.

\section{APPLICATIONS}
\begin{figure}
\plotone{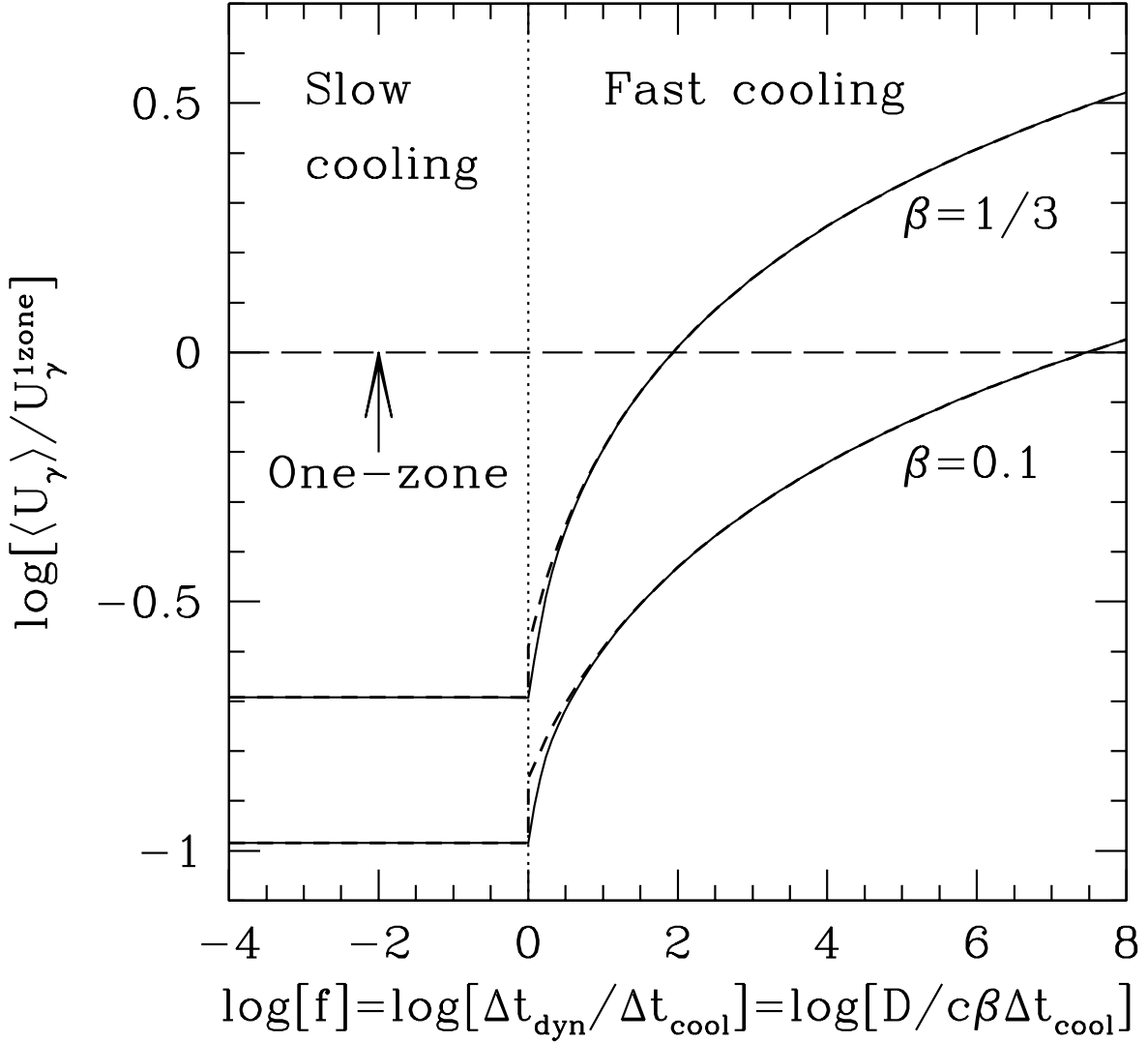}
\caption{The energy density of the seed photons divided
by the one-zone estimate $\langle U_{\gamma} \rangle/U_{\gamma}^{1zone}$
as a function of the ratio of the shell crossing time to the 
electron cooling time $f=\Delta t_{dyn}/\Delta t_{cool}=
D/c\beta \Delta t_{cool}$ is shown for the shock velocities
$\beta=1/3$ and $\beta=0.1$ with solid lines.
The analytical approximations in equations (\ref{eq:ugfast}) and
(\ref{eq:ugslow}) are also shown by dashed lines.
We call $f>1$ fast cooling and $f<1$ slow cooling.
}\label{fig:ave}
\end{figure}

(I) In the internal shocks of the GRBs 
(e.g., Piran 1999; Sari, Narayan \& Piran 1996),
the cooling time of electrons is about
$\Delta t_{cool}\sim 6\pi \gamma_e m_e c^2/\sigma_T c \gamma_e^2 B^2
\sim 10^{-6} \gamma_{e,3}^{-1} B_{6}^{2}\ {\rm s}$,
where $\gamma_e=10^3 \gamma_{e,3} \sim m_p/m_e$
is the Lorentz factor of the electrons
which contribute most of the radiation energy,
and $B=10^{6} B_{6}\ {\rm G}$ is the magnetic field.
The shock crossing time is about
$\Delta t_{dyn} \sim \Gamma d/c \sim 1 \Gamma_{2} d_{8} \ {\rm s}$,
where $\Gamma=10^2 \Gamma_{2}$ is the Lorentz factor of the shell
and $d=10^{8} d_{8} \ {\rm cm}$ is the shell thickness in the lab frame.
A large dispersion of the Lorentz factors
may be needed for the efficient internal shock
(Beloborodov 2000; Kobayashi \& Sari 2001).
Thus $\beta \sim 1/3$
and $f=\Delta t_{dyn}/\Delta t_{cool} \sim 10^{6}$.
From Fig.~\ref{fig:ave}, the energy density of the seed photons
is larger than the one-zone estimate by a factor of $\sim 3$.

(II) The late afterglows of the GRBs are likely in the slow cooling regime
(Sari \& Esin 2001).
Most of the radiation energy is emitted by electrons with $f=1$.
For the forward shock the shock velocity is $\beta\sim 1/3$,
while for the reverse shock the shock velocity has various values
depending on the parameters (Sari \& Piran 1995).
From Fig.~\ref{fig:ave}, the one-zone estimate
may overestimate the energy density of the seed photons
by order of unity for $\beta=0.1$.

(III) In the TeV blazars, most of the radiation is emitted
by electrons with the maximum Lorentz factor $\gamma_{max}\sim 10^5$
if the power law index of the electron distribution is $s<2$
(see Kino, Takahara \& Kusunose 2002).
Since electrons with the break Lorentz factor $\gamma_{br}\sim 10^3$
corresponds to the electrons with $f=1$ and the cooling time
is proportional to the inverse of the electron Lorentz factor,
the electrons with $\gamma_{max}$ have $f\sim 10^2$.
If we take the Klein-Nishina effect into account,
photons emitted by the electrons with $\gamma_{max}$ cannot be seed photons
for the electrons with $\gamma_{max}$.
In this case $f$ is reduced to $f\sim 10$ or so.
If the emission arises from the internal shock with the relative Lorentz factor
$\gamma_{r}\sim 2$ (Rees 1978), $\beta\sim 0.1$.
From Fig.~\ref{fig:ave}, the one-zone estimate
may overestimate the energy density of the seed photons
by a factor of $\sim 4$.

(IV) In the microquasars, according to Levinson \& Waxman (2001),
the ratio of the shock crossing time to the cooling time is 
about $f\sim 10^{5} l_{8}^{-1}$
where $l$ is the collision radius of the internal shock.
If we take the Klein-Nishina effect into account in the case $l_{8}=1$,
$f$ is reduced to $f\sim 10^{4}$ or so.
If the relative Lorentz factor between shells is $\gamma_{r}\sim 2$, 
$\beta\sim 0.1$.
From Fig.~\ref{fig:ave}, if $l=10^{8}$ cm, the one-zone estimate is 
correct within a factor of 2, while if $l=10^{13}$ cm,
the one-zone estimate
may overestimate the energy density of the seed photons
by order of unity.

\section{DISCUSSIONS}
We have not dealt with the emission spectrum
since we concern the total energy density in this paper.
If we consider the energy distribution of electrons,
the ultra fast cooling may modify the emission spectrum
conventionally used in the one-zone model.
This is because the cooling time depends on the electron energy,
i.e., less energetic electrons cool slower.
As expected from our analyses,
the seed photon field that electrons feel
will depend on the cooling time of the electrons.
Therefore the spectrum of the IC emission should be modified,
since the seed photon density is independent of the electron cooling time 
in the one-zone model.
Furthermore if the IC cooling dominates the synchrotron one,
the well known fact that the energy distribution
of the cooled electrons breaks by one power in the index
(e.g., Heavens \& Meisenheimer 1987)
should be modified because of the same reason.
Detail studies will be presented in the future paper.

To be precise, when the IC cooling dominates,
we have to solve the electron cooling
together with equation (\ref{eq:ugsum}).
This problem is the one-dimensional radiative transfer
with the electron cooling.
We may be able to solve this problem numerically,
although a large memory will be necessary for resolving 
the cooling layer in fast cooling.

We may have to consider other effects.
For example, the approximation of the uniform shell may not be good.
The reverse shock emission and the deceleration of the shock may be important.
If the synchrotron power is not constant but a function of the distance
from the center,
the logarithmic correction may become a power law one.
These are interesting future problems.

\acknowledgments
I am grateful F.~Takahara and M.~Kino for useful comments.
This work was supported in part by
Grant-in-Aid for Scientific Research 
of the Japanese Ministry of Education, Culture, Sports, Science
and Technology, No.00660.



\appendix

\section{INTEGRATION IN EQUATION (\ref{eq:ugsum})}\label{sec:app}

\begin{figure}
\plotone{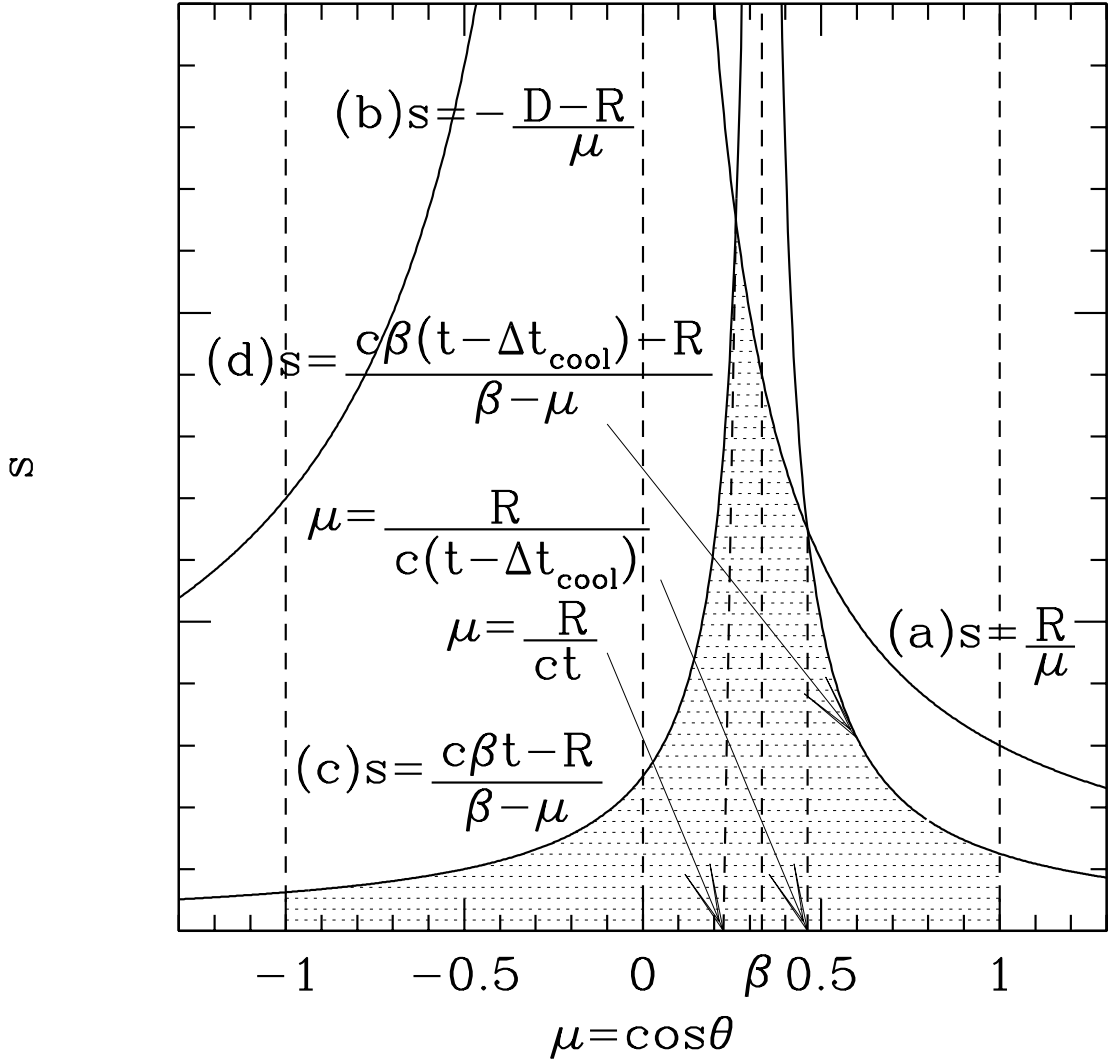}
\caption{The region to be integrated in equation (\ref{eq:ugsum})
is shaded in the $(\mu,s)$ plane.
Here we assume that equations (\ref{eq:temit}), (\ref{eq:t3}) 
and (\ref{eq:td}) are satisfied.
Four constraints on the integral are shown by solid lines,
(a) $s=R/\mu$, (b) $s=-(D-R)/\mu$,
(c) $s=(c\beta t-R)/(\beta-\mu)$
and (d) $s=[c\beta(t-\Delta t_{cool})-R]/(\beta-\mu)$.
The lines (a) and (c) cross at $\mu=R/ct$,
and the lines (a) and (d) cross at $\mu=R/c(t-\Delta t_{cool})$.
}\label{fig:region}
\end{figure}

In Fig.~\ref{fig:region}, the region to be integrated 
in equation (\ref{eq:ugsum}) is shaded.
There are four constraints on the integral in equation (\ref{eq:ugsum}),
which are shown by solid lines in Fig.~\ref{fig:region}.
(a) $s<R/\mu$ since the shell has an end $r=R-s\mu>0$
and we consider $R>0$.
(b) $s<-(D-R)/\mu$ since the shell has an end $r=R-s\mu<D$
and we consider $R<D$.
(c) $s<(c\beta t-R)/(\beta-\mu)$ since the first 
Heaviside step function in equation (\ref{eq:psyn})
has to be unity at the retarded time $t-s/c$, i.e.,
$t-s/c-(R-s\mu)/c\beta>0$.
(d) $s<[c\beta (t-\Delta t_{cool})-R]/(\beta-\mu)$ since the second
step function in equation (\ref{eq:psyn})
has to be zero at the retarded time $t-s/c$, i.e.,
$t-\Delta t_{cool}-s/c-(R-s\mu)/c\beta<0$.
Here we consider the interval during which electrons are emitting
in equation (\ref{eq:temit}),
so that the line (c) is in the region $\mu<\beta$
and the line (d) is in $\mu>\beta$ as in Fig.~\ref{fig:region}.
The lines (a) and (d) cross in the region $\beta<\mu<1$
as in Fig.~\ref{fig:region} if
equation (\ref{eq:t3}) is satisfied,
and vice versa.
The lines (b) and (c) do not cross in the region $-1<\mu<0$
as in Fig.~\ref{fig:region} if equation (\ref{eq:td}) is satisfied,
and vice versa.
If equations (\ref{eq:temit}), (\ref{eq:t3}) and (\ref{eq:td}) are 
satisfied,
equation (\ref{eq:ugsum}) can be integrated as
\beqa
U_{\gamma}(t,R)
&=&\frac{A}{2c}
\Biggl[\int^{1}_{R/c(t-\Delta t_{cool})}
d\mu \frac{c\beta(t-\Delta t_{cool})-R}{\beta-\mu}
\nonumber\\
&+&\int^{R/c(t-\Delta t_{cool})}_{R/ct}
d\mu \frac{R}{\mu}
+\int^{R/ct}_{-1} d\mu \frac{c\beta t-R}{\beta-\mu}\Biggr],
\label{eq:ugintA}
\eeqa
which gives equation (\ref{eq:ugint1}).
If equations (\ref{eq:temit}) and (\ref{eq:td}) 
are satisfied but equation (\ref{eq:t3}) is not,
we have
\beqa
U_{\gamma}(t,R)
=\frac{A}{2c}
\Biggl[\int^{1}_{R/ct}
d\mu \frac{R}{\mu}
+\int^{R/ct}_{-1} d\mu \frac{c\beta t-R}{\beta-\mu}\Biggr],
\label{eq:ugintB}
\eeqa
which gives equation (\ref{eq:ugint2}).

%
%

\end{document}